# The birth of a plasmonic topological quasiparticle on the nanofemto scale

Yanan Dai[1], Zhikang Zhou[1], Atreyie Ghosh[1], Roger S. K. Mong[1], Atsushi Kubo[2], Chen-Bin Huang[3], and Hrvoje Petek[1*]

[1]Department of Physics and Astronomy and Pittsburgh Quantum Institute, University of Pittsburgh, Pittsburgh, PA 15260, USA

[2]Division of Physics, Faculty of Pure and Applied Sciences, University of Tsukuba, 1-1-1 Tenno-dai, Tsukuba-shi, Ibaraki, 305-8571 Japan

[3]Institute of Photonics Technologies, National Tsing Hua University, Hsinchu 30013, Taiwan

**At interface of the classical and quantum physics Maxwell and Schrödinger equations describe how optical fields drive and control electronic phenomena at THz or PHz frequencies and on ultra-small scales to enable lightwave electronics.[1-5] Light striking a metal surface triggers electric field-electron particle/wave interactions to coherently imprint and transfer its attributes on the attosecond time scale. Here we create and image by ultrafast photoemission electron microscopy a new quasiparticle of optical field-collective electron interaction where the design of geometrical phase creates a plasmonic topological spin angular momentum texture. The spin texture resembles that of magnetic meron quasiparticle,[6] is localized within ½ wavelength of light, and exists on ~20 fs ($2\times10^{-14}$ s) time scale of the plasmonic field. The quasiparticle is created in a nanostructured silver film, which converts coherent linearly polarized light pulse into an evanescent surface plasmon polariton light-electron wave with a tailored geometric phase to form a plasmonic vortex. Ultrafast coherent microscopy imaging of electromagnetic waves propagating at the local speed of light of 255 nm/fs, electromagnetic simulations, and analytic theory find a new quasiparticle within the vortex core, with topological spin properties of a meron that are**

**defined by the optical field and sample geometry. The new quasiparticle is an ultrafast topological defect whose chiral field breaks the time-inversion symmetry on the nanoscale; its creation, symmetry breaking topology, and dynamics pertain to contexts ranging from the cosmological structure creation to topological phase transitions in quantum liquids and gases,[7-9] and may act as a transducer for quantum information on the nanofemto scale.[10,11]**

Classically, light as a wave propagates in the *k*-vector (momentum) direction with transverse electric ($E_L$) and magnetic ($H_L$) fields oscillating at a frequency $\omega$ and a phase $\varphi$; the field azimuthal direction and rotation defines its polarization and spin angular momentum (SAM). Quantum mechanically, it is also as a wave packet of photons, each carrying energy $\hbar\omega$ ($\hbar$ is the Planck's constant), momentum $\hbar k$, and SAM quantum $S=\pm 1\hbar$. When illuminating a solid, the $E_L$ field interacts with electric charges to generate collective light-matter quasiparticles such as excitons[12] and plasmons,[13] within the Fermi sea.[14,15] The quasiparticles are defined by the properties of the field, dielectric function of the solid, and geometry of the light-matter interaction. Upon capturing photons, quasiparticles can reemit them, decay into incoherent particles, or excite other degrees-of-freedom on the attosecond to femtosecond time scales.[16]

Light interacting with matter can also create electromagnetic topological textures as self-localized wave packets. Tony Skyrme proposed such quasiparticles to describe field textures in nucleons.[17] In magnetic materials, similar spin textures form through the Dzyaloshinskii–Moriya interaction with integer (skyrmion) or half-integer (meron) topological charges,[6,18] and have been found in Bose-Einstein condensates, superconductors, and ferroelectrics.[19] A plasmonic analog has been realized, where fields acquire skyrmion-like texture, through interference.[20]

Here, we create, image, and mathematically describe a new dynamical plasmonic quasiparticle by ultrafast photoelectron microscopy that is characterized by a time and field strength-dependent topological SAM texture like a magnetic meron, and transiently breaks the time-inversion symmetry. Light pulses illuminate a lithographically inscribed Archimedean spiral structure[21] in Ag film to generate surface plasmon polariton (SPP) wave packets with geometrically tailored phase fronts. SPPs are evanescent electromagnetic fields at metal/dielectric (e.g., Ag/vacuum) interfaces; as they propagate, their electric field ($E_{SPP}$) cycles into and out-of the surface plane, creating a transverse SAM pseudovector locked transversely to their $k$-vector,[22] a chiral property of SPPs known as the quantum spin Hall effect.[22-24] The geometric phase of the coupling structure[25] tailors SPPs with orbital angular momentum to steer them into a plasmonic vortex where their SAM acquires a dynamic meron-like texture (Fig. 1). We report how the orbital angular momentum and quantum spin Hall effect mold the plasmonic SAM into nanoscale meron-like texture on <20 fs time scale.[26] This new quasiparticle embodies the duration, phase, and polarization of structured light and can thus be applied to dynamical studies of topological phase transitions[7-9] or in quantum information processing.[10,11]

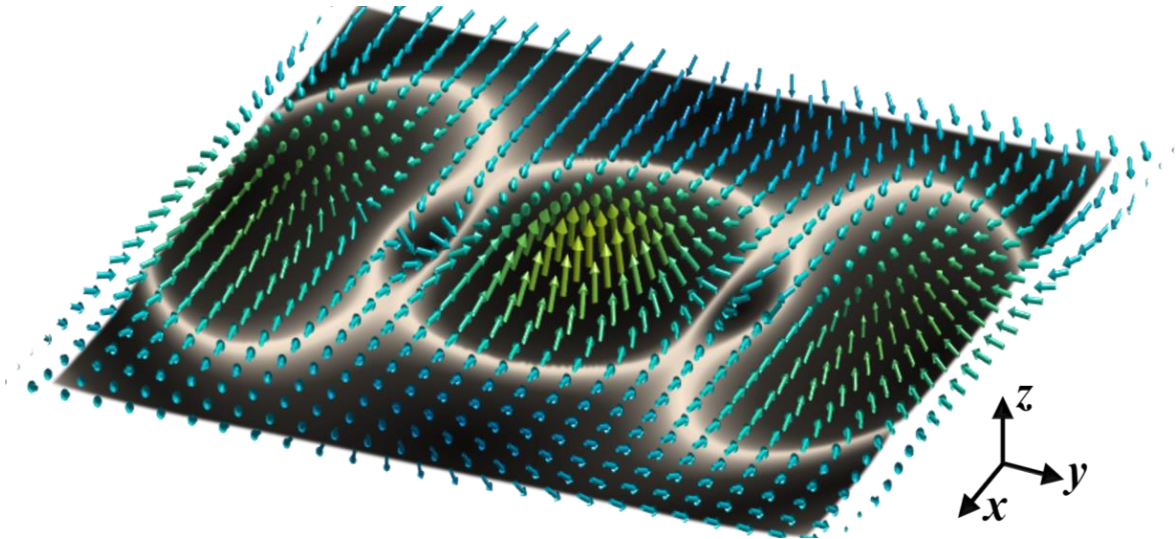

**Figure 1. The plasmonic meron SAM texture.** The arrows show meron-like SAM pseudovector texture at the plasmonic vortex core; they are overlaid on a map of L-line singularity of SPP fields that delineates the quasiparticle density.

To reveal the quasiparticle in Fig. 1, we capture the SPP flow and interference at 255 nm/fs by recording a movie with <10 nm spatial and ~100 as/frame pulse delay scan resolutions by interferometric time-resolved photoemission electron microscopy (ITR-PEEM).[27] In an ITR-PEEM experiment, a pair of identical, interferometrically scanned pump and probe light pulses illuminate the sample within the field-of-view of the microscope to excite the nonlinear two-photon photoemission (2PP) of electrons. PEEM images of the spatial distributions of emitted electron at each pump-probe pulse, which are proportional to the integral $\int E_T^4(x,y,z,t)dt$ of the total field at the sample (Fig. 2a). The total field, $E_T(x,y,z,t)=E_L(x,y,t)+E_{SPP}(x,y,z,t)$, is a superposition of the optical field and the SPP field it creates at the coupling structure; thus, PEEM records their spatially and temporally modulated interferences. Scanning the pulse delay and acquiring microscopic images records a movie of the fields and their interferences (for the experimental details and data see the Supplementary Information and the Movie 1). Movie 1 contains the dynamical information from interference between the in-plane (*x*,*y*) pump induced SPP wave and the probe optical fields that is modulated at $\omega_L$, and an unmodulated *z* component. The selective imaging of the dynamical in-plane component is accomplished by Fourier filtering the Movie 1 (see the Supplementary Information Section D and Movie 2). The PEEM imaging, electromagnetic simulations, and analytical model verify that illuminating the Archimedean structure with a geometric charge *m*=2, creates a plasmonic vortex[28] with a meron-like SAM texture at its core. The experiment and analysis confirm a novel coherent quasiparticle that exists dynamically as a condensed matter topological defect that breaks the time-inversion symmetry.

By illuminating the broken Archimedean coupling structure in Fig. 2b with *y*-linearly polarized *λ*=550 nm light, the oppositely propagating SPP waves emanating from each side carry its geometric charge to form a plasmonic vortex at its center. Figure 2c shows a single frame extracted from the ITR-PEEM Movie 1, which reveal the vortex, where SPP fields interfere causing them to gyrate about a phase singularity at its core.

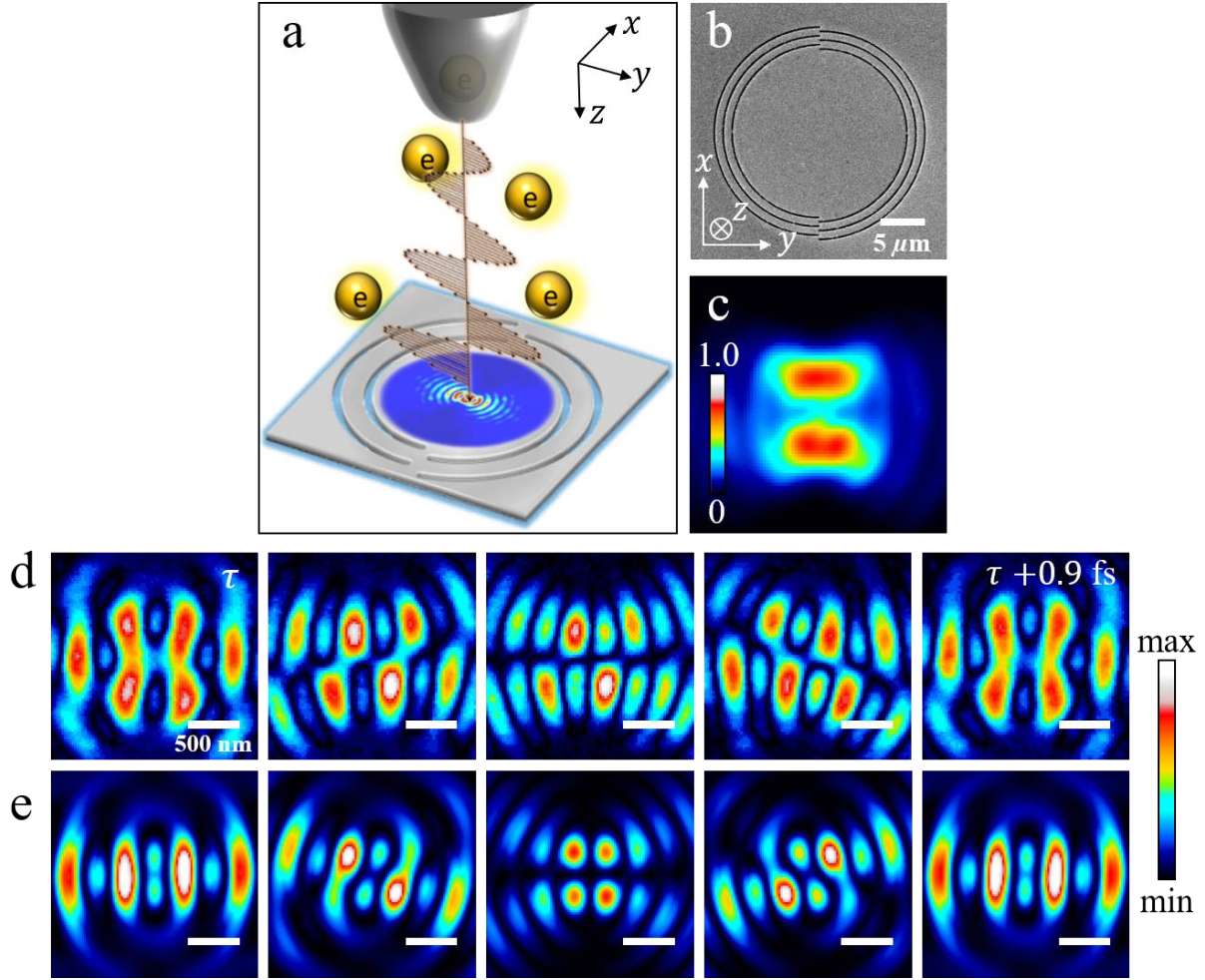

**Figure 2. Ultrafast microscopy of the SPP vortex. a,** Schematic of the experimental setup for the vortex generation and imaging. **b,** Scanning electron micrograph of the SPP vortex generator with a geometric charge of *m*=2 etched in an Ag film. **c,** Static PEEM image of the SPPs vortex excited by *y*-polarized light. **d,** Time-resolved, Fourier filtered (*1ω$_L$*) PEEM images (absolute value), which reveal the gyrating interferences of SPP fields; advancing the delay by *Δt*~1 fs, or π optical cycle, causes the interference fringe phases to advance superluminally by 510 nm/fs in one-half of an optical cycle. Evolution of two constructive fringes is marked by the white arrows and black dots, respectively. **e,** The FDTD calculated photoemission distributions corresponding to the PEEM images in (d).

Figure 2d reveals the vortex phase evolution (rotation) from the Fourier filtered Movie 2 where for *i-v*, $\Delta$t is advanced by ~1 fs, or a phase delay of π, causing the interference fringes to flow on top and bottom out-of and into alignment in a clockwise circulating motion. We perform finite difference time domain (FDTD) simulation of the vortex generation and imaging; the simulated PEEM images (Fig. 2e) reproduce the experimental fringe circulation.

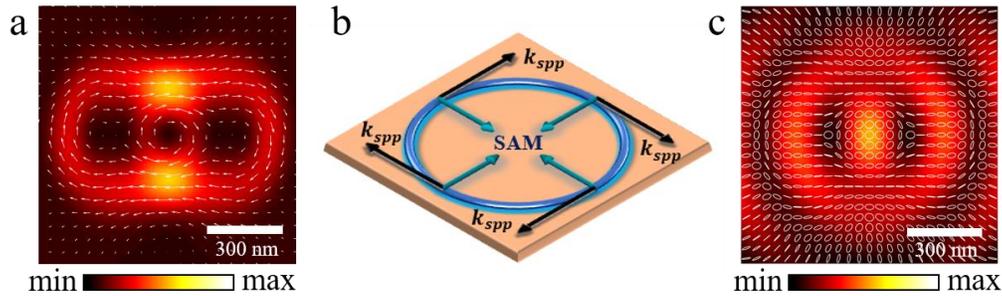

**Figure 3. The origin of the topological SAM texture. a,** Poynting vectors showing the flow of SPP energy magnitude (color scale) and direction (arrows) of the clockwise vortex. **b,** Parallel to the Poynting vectors, the orbiting of the SPP *k*-vector about the vortex has a transverse, inward pointing SAM. **c,** Amplitude of the in-plane $E_{SPP}$ field (color scale; equivalent to the first Stokes parameter, $P_0$), and its in-plane polarization states (the overlaid ellipses); the center circular polarization generates the out-of-plane SAM.

Figure 3 shows the spatial vectorial properties of SPP vortex fields. The Poynting vectors in Fig. 3a, which define the *k*-vector distribution, whose clockwise circulation causes SAM to point into the vortex center as a consequence of spin-momentum locking (Fig. 3b). Figure 3c plots the calculated in-plane $E_{SPP}$ field vectors, which at the vortex center circulate clockwise, but farther away, pass through a linearly polarized fringes. In vectorial optics, these limiting polarizations are named circular C-point and linear L-line singularities;[29] they define the SAM texture in Fig. 1. We further determine an L-line map (detailed in Supplementary Information section F and Movie 3), where the SAM passes through the *x,y*-plane (white shading in Fig. 1 and 4a), by calculating the eccentricity distribution of the $E_{SPP}(x,y)$ fields and its time evolution. The same map is calculated analytically, or by locating interfaces where the *z*-component of SAM changes sign (Fig. 4b). Specifically, the central L-line hourglass-shaped contour defines the meron texture edge where the SAM points in-plane into the vortex, undergoing $2\pi$ rotation. The vortex and meron centers coincide at the C-point, where the circulating in-plane $E_{SPP}$ field, by Faraday's law, generates a surface normal *H* field, breaking the time-inversion symmetry and causing SAM to point in the positive *z*-direction, as seen in Fig. 1.

Next, we address the topological charge and stability of the meron-like quasiparticle by evaluating the SPP spin angular momentum ***S***,

$$\boldsymbol{S} = \frac{1}{2\omega} Im(\varepsilon \boldsymbol{E}^* \times \boldsymbol{E} + \mu \boldsymbol{H}^* \times \boldsymbol{H}), \tag{1}$$

and the related topological particle density, *D*,

$$D = \frac{1}{4\pi} \boldsymbol{S}' \cdot \left(\frac{\partial \boldsymbol{S}'}{\partial x} \times \frac{\partial \boldsymbol{S}'}{\partial y}\right). \tag{2}$$

where ***S'*** is the SAM normalized to 1. The integral of *D* within the boundary L-line region gives the topological charge, which is quantized in multiples of ½. *D* evolves in time (Movie 4 and Supplementary Information) to reach a steady distribution, which we plot in Fig. 4c.

Using the central hourglass-shaped L-line as the quasiparticle boundary, and integrating *D*

within it gives a topological charge of Q=½ that defines a stable meron texture while the vortex exists.[6] Moreover, the continuous rotation of the edge SAM by $2\pi$ (Fig. 4e) along the L-line hourglass boundary, and its continuous change to central *z*-orientation, further confirm the meron texture. Figure 4f shows that even though $E_{SPP}$ oscillates at the laser driving frequency, the topological charge is constant as long as the vortex is defined; this analysis shows that an excitation geometry dependent meron-like quasiparticle forms with ~$\lambda/2$ dimension that breaks time-inversion symmetry on time scale of the generating optical field. We confirm the SAM stability over >10 optical cycles (~20 fs) by plotting the plasmonic flow[30] in Fig. 4d, which reproduces the L-line map and the hourglass boundary (Supplementary Information section I).

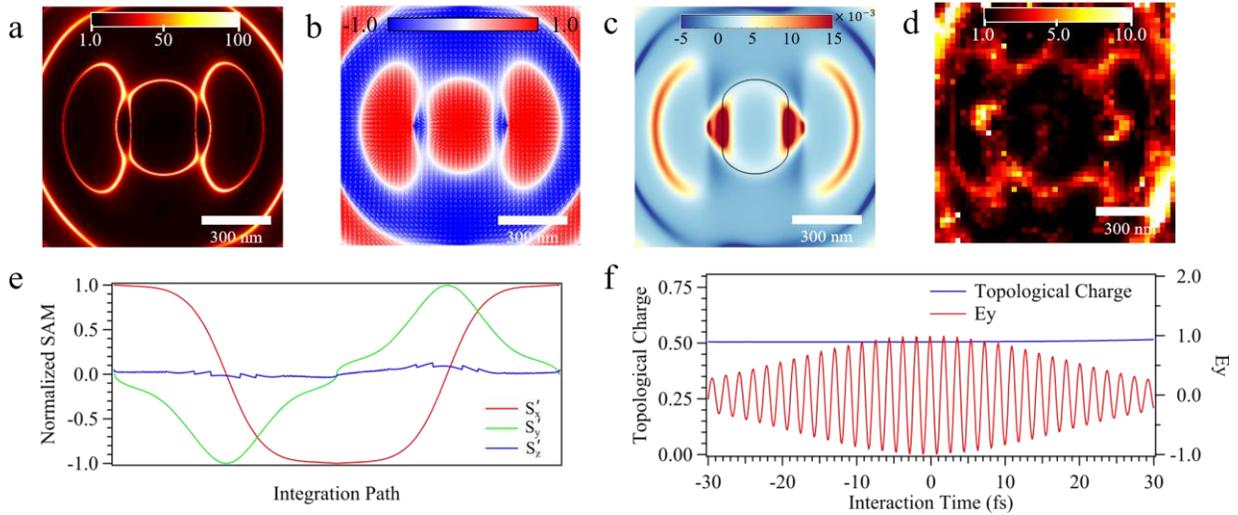

**Figure 4. The topological charge. a,** The L-line singularity map of the local in-plane polarization of SPP fields near the vortex core. The central hourglass shaped L-line contour encloses the meron SAM density. **b,** Distribution of the z-component (color scale) of the normalized SAM vectors, where white interfaces correspond to the in-plane polarization; the arrows plot the in-plane components. **c,** The quasiparticle density *D* map for the *m*=2 plasmonic vortex from Eq. (2), when the SPP field is maximum at the vortex. The L-line hourglass (black

line) indicates the integration area to obtain the topological charge. **d,** The experimental L-line map from the photoemission intensity flow over 10 optical cycles, which localizes the meron and confirms its stability. **e,** The normalized SAM vector components taken along the hourglass L-line. **f,** The topological charge of Q=½ and the 2π rotation of the in-plane SAM in (e) define the meron texture; it is stable while the vortex exists (blue), whereas the *y* component of $E_{SSP}$ at the vortex core (red) oscillates at $\omega_L$.

The generated quasiparticle enables nanofemto control of SAM topological textures with lifetime and amplitude defined by the generating optical field. Different choices of the geometrical coupling structure and light polarization define the topological charge, and therefore, whether the generated texture is meron-like, skyrmion-like, or an array of interacting topological quasiparticles.[31] The quasiparticle field is sufficiently intense to drive the nonlinear two-photon photoemission that we detect. We note that vortices occur in optical speckle,[10,29] and thus such topological defects can emerge spontaneously, for example, in a cosmological context, to drive structure formation.[8,9] In optical applications, pulses of different durations, field strengths, polarizations, or with actively manipulated phase and amplitude, will enable the creation, manipulation, and annihilation of plasmonic topological spin textures on a few optical cycle time scales (<10 fs). The generated quasiparticles can transfer their spin textures to proximate quantum gasses (Bose-Einstein condensates and superconductors), modulate spins of defect centers, imprint topological properties on other matter, and by breaking the time-inversion symmetry, drive gyroelectric[32] effects such as rotation of polarization of optical fields.[33] The tailored SAM textures can interact with other degrees-of-freedom, such as ballistic electrons at the Fermi level,[6] to exchange energy, momentum, and spin among them to enable optical switching and quantum state control on a few femtosecond and nanometer scales.

**Acknowledgements.** This research was supported by the NSF Center for Chemical Innovation on Chemistry at the Space-Time Limit grant CHE-1414466. The Authors thank Prof. Luat Vuong for advice on the plasmonic flow analysis. The authors thank the Peterson Institute of NanoScience and Engineering for help in the sample preparation.

**Author contributions.** YD performed the experiments and simulations, processed and analyzed the data, and generated the figures; ZZ helped with the sample preparation and performing the experiments; YD, ZZ and AG developed the analytical model for topological quasiparticles; RSKM advised on the topological properties of quasiparticles; AK explained the PEEM imaging of plasmonic phenomena; C-BH introduced the research on plasmonic vortices, performed simulations of experiments, and participated in the data interpretation, HP introduced the concept of topological quasiparticles, wrote the manuscript, and supervised the research; all authors contributed to the discussion.

**Competing interests.** The authors declare no competing interests.

**Additional information**

**Correspondence.** should be addressed to HP.

**Supplementary information** is available for this paper at